\documentclass[aps,prd,showpacs,twocolumn,floatfix,amsmath,amssymb,superscriptaddress]{revtex4-1}
\usepackage{hyperref}
\usepackage{bm}
\usepackage{graphicx}

\newcommand{\beq}{\begin{equation}}
\newcommand{\eeq}{\end{equation}}
\newcommand{\beqa}{\begin{eqnarray}}
\newcommand{\eeqa}{\end{eqnarray}}
\newcommand{\deltam}{\delta_\mathrm{m}}
\newcommand{\Omegam}{\Omega_\mathrm{m}}
\newcommand{\GW}{\mathrm{GW}}
\newcommand{\Mpc}{\mathrm{Mpc}}
\newcommand{\Gpc}{\mathrm{Gpc}}
\newcommand{\hMpc}{h^{-1}\,\mathrm{Mpc}}

\newcommand{\mnras}{Mon. Not. Roy. Astron. Soc.}
\newcommand{\apjl}{Astrophys. J. Lett.}

\newcommand{\physrep}{Phys. Rept.}
\newcommand{\aap}{Astron. Astrophys.}

\newcommand{\ssr}{Space Sci. Rev.}

\newcommand{\pasj}{Publ. Astron. Soc. Jpn.}

\begin{document}

\title{Exploring the distance-redshift relation
with gravitational wave standard sirens and tomographic weak lensing}

\author{Ken Osato}
\email[]{ken.osato@utap.phys.s.u-tokyo.ac.jp}
\affiliation{Department of Physics, School of Science, The University of Tokyo,
113-0033 Tokyo, Japan}

\date{\today}

\begin{abstract}
Gravitational waves from inspiraling compact objects provide us with
information of the distance scale
since we can infer the absolute luminosity of the source from analysis of the wave form,
which is known as standard sirens.
The first detection of the gravitational wave signal of
the binary black hole merger event by Advanced LIGO has opened up
the possibility of utilizing standard sirens as cosmological probe.
In order to extract information of the distance-redshift relation,
we cross-correlate weak lensing, which is an unbiased tracer of matter distribution
in the Universe, with the projected number density of gravitational wave sources.
For weak lensing, we employ tomography technique to efficiently obtain
information of large-scale structures at wide ranges of redshifts.
Making use of the cross-correlations along with the auto-correlations,
we present forecast of constraints on four cosmological parameters,
i.e., Hubble parameter, matter density, the equation of state parameter of dark energy,
and the amplitude of matter fluctuation.
To fully explore the ability of cross-correlations, which require large overlapping sky coverage,
we consider the specific case with
the upcoming surveys by \textit{Euclid} for weak lensing
and Einstein Telescope for standard sirens.
We show that cosmological parameters can be tightly constrained solely by
these auto- and cross-correlations of standard sirens and weak lensing.
For example, the $1\text{-}\sigma$ error of Hubble parameter is expected to be
$\sigma (H_0) = 0.33 \, \mathrm{km} \, \mathrm{s}^{-1} \, \mathrm{Mpc}^{-1}$.
Thus, the proposed statistics will be a promising probe into the distance scale.
\end{abstract}


\maketitle

\section{Introduction}
\label{sec:intro}
The first detection of gravitational wave (GW) signal from merging binary black holes (BH),
GW150914, by Advanced Laser Interferometer Gravitational Wave Observatory (LIGO)
provides us with the new probe into cosmology and astrophysics \cite{Abbott16a,Abbott16b}.
After four successful detections of GW signals from black hole mergers
(GW151226 \cite{Abbott16c}, GW170104 \cite{Abbott17a}, GW170608 \cite{Abbott17b}, and GW170814 \cite{Abbott17c}),
the first detection of the GW signal from a neutron star (NS) binary is reported (GW170817) \cite{Abbott17d}.
Aiming for detection of more sources and better localization,
several projects of interferometers have been proposed;
Advanced Virgo \cite{Acernese15} has started observing run,
KAGRA \cite{Somiya12} is on commissioning,
and LIGO-India \cite{Unnikrishnan13} has been approved for construction.

Once this network is established, it enables us to search the GW sources
for the whole sky with high sensitivity.
Furthermore, more telescopes both on ground and in space, e.g.,
Einstein Telescope \cite{Punturo10}, Cosmic Explorer \cite{Abbott17e},
LISA \cite{Amaro-Seoane12,Amaro-Seoane13},
and DECIGO \cite{Kawamura08}, are planned to achieve
unprecedented measurements of GW signals over wide ranges of frequency.
These telescopes will enable us to detect
large numbers of GW sources with accurate wave forms.

One of the important aspects of GW measurements is that from the observed wave form
we can measure the amplitudes both at observer and source frames.
Thus, we can infer the luminosity distance of the source (\textit{standard sirens}).
If the redshift of the GW source is known, we can investigate the geometry
of the Universe through the \textit{distance-redshift relation}
(see, e.g., Ref.~\cite{Seto18}).
However, solely with GW observations, inferring the redshift of the source
is quite demanding.
One of methods to estimate the source redshift is
to observe electro-magnetic (EM) counterpart of the GW event.
For the NS binary event GW170817, the EM counterpart has been
detected with optical imaging observation \cite{Utsumi17,Tanaka17,Tominaga18},
but detecting a counterpart is still challenging due to the short time scale of GW events and
large uncertainty of localization with current interferometers.
On the other hand, without redshift information, the anisotropic distribution of
GW sources can be used as cosmological probe \cite{Namikawa16}.
Similarly to the number density distribution of galaxies, we can naively expect that
the number density of compact object binaries should reflect
the large scale matter density distribution.
Accordingly, statistics of GW source distribution such as two-point correlation functions
can be used to probe into cosmology.

Though the GW source distribution itself is useful for cosmology,
when combining another observable which redshift information is accessible,
we can investigate the distance-redshift relation indirectly.
One of such candidates is the spatial distribution of
spectroscopically observed galaxies \cite{Oguri16},
since the redshift of such galaxies are precisely determined.
However, there is a drawback of using the spectroscopic galaxy samples.
In order to obtain cosmological information,
we need to introduce a galaxy bias which relates the galaxy number density distribution
with matter fluctuation. Practically, the bias is treated as a free parameter
and marginalized finally. This degrades the constraints on cosmological parameters.
For better parameter determination, we need
another cosmological probe, in which redshift information is available
and robust to systematics.
In this work, we focus on weak gravitational lensing (WL).
One of advantages is that WL is an unbiased tracer of density fluctuation,
which does not necessitate a bias parameter.
However, since the observables of WL is a projected quantity,
information of matter distributions at different redshifts are entangled.
We can evade this problem with technique known as \textit{tomography} \cite{Hu99}.
The whole source galaxy samples can be divided according to
photometric redshifts of source galaxies.
Then, one can construct observables of WL using galaxies in each redshift bin,
and measure auto- and cross-correlations of observables.
As a result, we can efficiently obtain information of matter distribution at various redshifts.

Recently, various works are devoted to probing the distance-redshift relation
utilizing standard sirens, e.g., auto-correlation of
GW source distribution \cite{Namikawa16} and
cross-correlation between standard sirens and galaxy distributions \cite{Oguri16}.
In this paper, we address the cross-correlation between tomographic weak lensing
and GW source distributions. Similarly to the measurement of galaxy clustering,
forthcoming weak lensing surveys cover large areas.
Therefore, combining these measurements has a possibility
to place a very tight constraint on cosmological models.

This paper is organized as follows. First, we give formulation of auto- and
cross-correlations of tomographic weak gravitational lensing
and source distribution of GW signals.
Then, we forecast how cosmological parameters can be constrained with
upcoming GW and weak lensing measurements. We adopt flat $\Lambda$ cold dark matter model,
and cosmological parameters;
Hubble parameter $H_0 = 100 h \, \mathrm{km} \, \mathrm{s}^{-1} \, \Mpc^{-1} =
67.27 \, \mathrm{km} \, \mathrm{s}^{-1} \, \Mpc^{-1}$, the present day density parameters of
cold dark matter and baryon $\Omega_\mathrm{c} h^2  = 0.1198$, $\Omega_\mathrm{b} h^2 = 0.2225$,
the tilt and the amplitude of the scalar perturbation
$n_\mathrm{s} = 0.9645$, $A_\mathrm{s} = 2.2065 \times 10^{-9}$,
at the pivot scale $k_\mathrm{piv} = 0.05 \, \mathrm{Mpc}^{-1}$, and
the total mass of neutrinos $M_\nu = 0.06 \, \mathrm{eV}$
based on the measurements of the anistropy of temperature and polarization of
cosmic microwave background (TT, TE, EE+lowP) by the \textit{Planck} mission \cite{Planck16}.
There are derived parameters which will be used later;
the total matter density parameter $\Omega_\mathrm{m} =
\Omega_\mathrm{c}+\Omega_\mathrm{b} = 0.3153$,
and the amplitude of matter fluctuation at the scale of $8 \, \hMpc$,
$\sigma_8 = 0.831$.
We assume that the neutrino component consists of two massless and one massive neutrinos.

\section{Formulation}
\label{sec:formulation}
In this section, we formulate how one can compute the auto- and cross-correlations
of the GW source number density and WL convergence field.

\subsection{Gravitational wave sources}
In the measurements of merging binaries of compact objects,
the luminosity distances can be obtained from the wave form.
However, the estimated luminosity distance can deviate from the true value
due to several uncertainties, e.g., degeneracy with other parameters
such as the mass of the compact objects or the inclination angle,
and statistical fluctuation.
We assume that the inferred luminosity distance $\hat{D}$ follows
the log-normal distribution where the mean is the true one $D$,
\beq
p(\hat{D} | D) = \frac{1}{\sqrt{2 \pi} \sigma_{\ln D} \hat{D}}
\exp [-x^2 (\hat{D}, D)] ,
\eeq
where
\beq
x(\hat{D}, D) \equiv \frac{\ln \hat{D} - \ln D}{\sqrt{2} \sigma_{\ln D}} ,
\label{eq:xD}
\eeq
and we adopt $\sigma_{\ln D} = 0.05$.
In addition, the estimate of the luminosity distance is subject to
weak gravitational lensing by intervening matter in the Universe.
Since the object looks brighter due to the magnification effect,
the luminosity distance becomes smaller compared with the case of no lensing.
This effect can be expressed as,
\beq
D = \bar{D}(z) \mu^{-\frac{1}{2}}(\bm{\theta}, z)
\simeq \bar{D}(z) [1-\kappa(\bm{\theta}, z) ] ,
\eeq
where $\bar{D}$ is the luminosity distance computed
in the flat Friedmann-Lema\^itre-Robertson-Walker metric.
In the weak field limit, the magnification $\mu$ is approximated as $1+2\kappa$,
where $\kappa$ is the convergence field.
The convergence corresponds to the projected matter density contrast $\deltam$
convolved with distance kernel,
\beqa
\kappa (\bm{\theta}, \chi) &=& \frac{3 H_0^2 \Omegam}{2 c^2}
\int_0^\chi d\chi' \, \frac{\chi' (\chi - \chi')}{\chi}
\frac{\deltam(\chi' \bm{\theta}, \chi')}{a(\chi')} \nonumber \\
&\equiv& \int_0^\chi d\chi' \, W^\kappa (\chi; \chi')
\deltam(\chi' \bm{\theta}, \chi'),
\label{eq:kappa}
\eeqa
where $\chi$ is comoving distance from the observer and
$a$ is the scale factor.
Hereafter, we adopt the comoving distance as the indicator of the cosmic time
instead of the redshift. However, we can convert each other by the relation,
\beq
\chi (z) = \int_0^z \frac{c dz'}{H(z')} .
\eeq

Then, let us consider the number density field of GW sources.
We divide the whole sources according to the observed luminosity distance.
For $i$th bin we select sources with
$D_{i, \mathrm{min}} < \hat{D} < D_{i, \mathrm{max}}$.
The number density field is obtained by projecting sources as
\beq
n^{\mathrm{w}}_i (\bm{\theta}) = \int_0^{\chi_\mathrm{H}} \! d\chi \,
\chi^2 G_i (\chi) n_\GW (\chi \bm{\theta}, \chi),
\eeq
where $\chi_\mathrm{H}$ is the comoving distance to the horizon,
$G_i (\chi, \bm{\theta})$ is the selection function,
\beqa
G_i (\chi, \bm{\theta}) \equiv \frac{1}{2}
[ \mathrm{erfc} \{ x(D_{i, \mathrm{min}}, D(\chi, \bm{\theta}) \}
\nonumber \\
-\mathrm{erfc} \{ x(D_{i, \mathrm{max}}, D(\chi, \bm{\theta}) \} ] ,
\eeqa
and $n_\GW$ is the three-dimensional number density of GW sources.
Since the modulation effect on the luminosity distance due to lensing
is relatively small, one can Taylor expand the selection function as
\beqa
G_i (\chi, \bm{\theta}) &\simeq& G_i |_{D=\bar{D}} +
\left. \frac{d G_i}{d D} \right|_{D=\bar{D}} (D-\bar{D}) \nonumber \\
&=& \frac{1}{2} [ \mathrm{erfc} \{ x(D_{i, \mathrm{min}}, \bar{D} (\chi) \}
-\mathrm{erfc} \{ x(D_{i, \mathrm{max}}, \bar{D} (\chi) \}] \nonumber \\
&& + \kappa(\chi \bm{\theta}, \chi)
\frac{1}{\sqrt{2 \pi} \sigma_{\ln \mathrm{D}} }
\{ -\exp [-x^2 (D_{i, \mathrm{min}}, \bar{D} (\chi))] \nonumber \\
&& +\exp [-x^2 (D_{i, \mathrm{max}}), \bar{D} (\chi)] \} \nonumber \\
&\equiv& S_i (\chi) + \kappa(\chi \bm{\theta}, \chi) T_i (\chi) .
\label{eq:Gterm}
\eeqa
The averaged number density is expressed as
\beqa
\bar{n}^\mathrm{w}_i &=& \int_0^{\chi_\mathrm{H}} \! d\chi \,
\chi^2 S_i (\chi) \bar{n}_\GW (\chi) \nonumber \\
&=& \int_0^{\chi_\mathrm{H}} \! d\chi \,
\chi^2 S_i (\chi) T_\mathrm{obs} a(\chi) \dot{n}_\GW (\chi) ,
\label{eq:meandensity}
\eeqa
where $T_\mathrm{obs}$ is the duration of the observation
and $\dot{n}_\GW (\chi)$ is the rate density of detectable merger events.
Since the convergence vanishes when averaged in angular space,
only the first term in Eq.~(\ref{eq:Gterm}) remains.

We can construct the two-dimensional number density contrast of GW sources as
\beqa
\delta^\mathrm{w}_i (\bm{\theta}) &\equiv&
\frac{n^\mathrm{w}_i (\bm{\theta}) - \bar{n}^\mathrm{w}_i}
{\bar{n}^\mathrm{w}_i} \nonumber \\
&=& \frac{1}{\bar{n}^\mathrm{w}_i} \int_0^{\chi_\mathrm{H}} \! d\chi \,
\chi^2 S_i (\chi) \bar{n}_\GW (\chi) \delta_\GW (\chi \bm{\theta}, \chi) \nonumber \\
&& + \frac{1}{\bar{n}^\mathrm{w}_i} \int_0^{\chi_\mathrm{H}} \! d\chi \,
\chi^2 T_i (\chi) \bar{n}_\GW (\chi) \kappa (\chi \bm{\theta}, \chi) .
\eeqa
We can rewrite the second term and define a kernel as,
\begin{widetext}
\beqa
&& \frac{1}{\bar{n}^\mathrm{w}_i} \int_0^{\chi_\mathrm{H}} \! d\chi \,
\chi^2 T_i (\chi) \bar{n}_\GW (\chi) \kappa (\chi \bm{\theta}, \chi)
= \frac{1}{\bar{n}^\mathrm{w}_i} \int_0^{\chi_\mathrm{H}} \! d\chi \int_0^\chi d\chi' \,
\chi^2 T_i (\chi) \bar{n}_\GW (\chi) W^\kappa (\chi; \chi')
\deltam(\chi' \bm{\theta}, \chi') \nonumber \\
&=& \int_0^{\chi_\mathrm{H}} d\chi'
\left( \frac{1}{\bar{n}^\mathrm{w}_i} \int_{\chi'}^{\chi_\mathrm{H}} \! d\chi \,
\chi^2 T_i (\chi) \bar{n}_\GW (\chi) W^\kappa (\chi; \chi') \right)
\deltam(\chi' \bm{\theta}, \chi')
\equiv \int_0^{\chi_\mathrm{H}} \! d\chi' \, W^{\mathrm{t}}_i (\chi')
\deltam(\chi' \bm{\theta}, \chi') .
\label{eq:Wtkernel}
\eeqa
\end{widetext}
Similarly, we also define the kernel in the first term,
\beqa
&& \frac{1}{\bar{n}^\mathrm{w}_i} \int_0^{\chi_\mathrm{H}} \! d\chi \,
\chi^2 S_i (\chi) \bar{n}_\GW (\chi)
\delta_\GW (\chi \bm{\theta}, \chi) \nonumber \\
&=& \int_0^{\chi_\mathrm{H}} \! d\chi
\left( \frac{1}{\bar{n}^\mathrm{w}_i} \chi^2 S_i (\chi) \bar{n}_\GW (\chi)
b_\GW \right) \deltam (\chi \bm{\theta}, \chi) \nonumber \\
&\equiv& \int_0^{\chi_\mathrm{H}} \! d\chi \,
W^\mathrm{s}_i (\chi) \deltam(\chi \bm{\theta}, \chi) .
\label{eq:Wskernel}
\eeqa
Here we assume the linear bias relation $\delta_\GW = b_\GW \deltam$
and the bias is absorbed in the kernel $W^\mathrm{s}_i$.

\subsection{Tomographic weak lensing}
WL has now been measured by optical surveys and enables one to
constrain cosmological models (for comprehensive reviews,
see Refs.~\cite{Bartelmann01,Kilbinger15}).
It gives rich information about the large-scale structures in the Universe.
WL is characterized by convergence $\kappa$ and shears $\gamma_1$ and $\gamma_2$.
It is possible to transform the convergence into shears and vice versa.
In this paper, we focus only on the convergence field.
As is shown in Eq.~(\ref{eq:kappa}), the convergence can be described as the projection of
the matter density field, but in real surveys, the redshift distribution
of source galaxies has a broad shape.
Then, the observable is the one convolved with the source distribution,
\beqa
\kappa^\mathrm{G}_i (\bm{\theta}) &=& \int_0^{\chi_\mathrm{H}} \! d\chi \, p_i (\chi)
\kappa (\bm{\theta}, \chi) \nonumber \\
&=& \int_0^{\chi_\mathrm{H}} \! d\chi \, W^\mathrm{G}_i (\chi)
\deltam(\chi \bm{\theta}, \chi),
\eeqa
where $p_i (\chi)$ is the comoving distance distribution of source galaxies,
and the kernel is given as
\beqa
W^\mathrm{G}_i (\chi) &\equiv&
\int_\chi^{\chi_\mathrm{H}} \! d\chi' \, p_i (\chi') W^\kappa (\chi'; \chi)
\nonumber \\
&=& \frac{3 H_0^2 \Omegam}{2 c^2}
\int_\chi^{\chi_\mathrm{H}} \! d\chi' \, \frac{p_i (\chi')}{a(\chi)}
\frac{\chi (\chi' - \chi)}{\chi'} .
\label{eq:WGkernel}
\eeqa
This distribution is normalized as unity, i.e.,
\beq
\int_0^{\chi_\mathrm{H}} \! d\chi \, p_i (\chi) = 1 .
\label{eq:pnorm}
\eeq
The subscript $i$ represents the label of the source samples.
According to the photometric redshifts of the source galaxies,
we can divide the whole sample with different redshift distributions.
Thus, we can probe the evolution of structures.
This technique is called as lensing tomography \cite{Hu99}.

In addition to weak lensing effect,
the shape of the galaxy is subject to the local tidal field.
Since this tidal field is correlated with the large-scale structure as well,
it modulates the observed convergence field.
This effect is referred to as intrinsic alignment (IA)
(for reviews, see Refs.~\cite{Troxel15,Joachimi15}).
We quantify this effect based on nonlinear-linear alignment model
\cite{Hirata04,Bridle07,Joachimi11},
\beqa
\kappa^\mathrm{I}_i (\bm{\theta}) &=& \int_0^{\chi_\mathrm{H}} \! d\chi \,
p_i (\chi) \left( - A_\mathrm{IA} C_1 \rho_\mathrm{cr}
\frac{\Omega_\mathrm{m}}{D_+(\chi)} \right)
\deltam(\chi \bm{\theta}, \chi) \nonumber \\
&\equiv& \int_0^{\chi_\mathrm{H}} \! d\chi \,
W^\mathrm{I}_i (\chi) \deltam(\chi \bm{\theta}, \chi) ,
\label{eq:WIkernel}
\eeqa
where $\rho_\mathrm{cr}$ is the critical density, $D_+(\chi)$ is
the linear growth factor which is normalized to unity at present,
$A_\mathrm{IA}$ is a free parameter which determines the amplitude and
$C_1 = 5 \times 10^{-14} \, h^{-2} \, \mathrm{M}_\odot^{-1} \, \Mpc^3$.
This model has been to applied to real data (see, e.g., Ref.~\cite{Hildebrandt17}),
and the dependence of the amplitude on redshift and source luminosity
is shown to be very weak \cite{Joudaki17}.
As a result, the convergence field is observed as the sum of two contributions,
\beq
\kappa_i = \kappa^\mathrm{G}_i + \kappa^\mathrm{I}_i .
\eeq

\subsection{Auto- and cross-power spectra}
Here, we construct power spectra of the GW source number density and WL.
The angular power spectra are defined as,
\beq
\langle X_{\ell m} Y^*_{\ell' m'} \rangle \equiv
\delta_{\ell \ell'} \delta_{m m'} C_{XY}(\ell) ,
\eeq
where $X_{\ell m}$ and $Y_{\ell m}$ are the coefficient of
spherical harmonic expansion of either $\delta^\mathrm{w}_i$ or $\kappa_i$,
and the parenthesis denotes ensemble average.
The auto-spectra of GW source number density $C_{\mathrm{w}_i \mathrm{w}_j}$
and convergence $C_{\mathrm{l}_i \mathrm{l}_j}$ and
their cross-spectra $C_{\mathrm{w}_i \mathrm{l}_j}$ are given as
\beqa
C_{\mathrm{w}_i \mathrm{w}_j} (\ell) &=&
C_{\mathrm{s}_i \mathrm{s}_j} + C_{\mathrm{s}_i \mathrm{t}_j} +
C_{\mathrm{t}_i \mathrm{s}_j} + C_{\mathrm{t}_i \mathrm{t}_j} , \\
C_{\mathrm{l}_i \mathrm{l}_j} (\ell) &=&
C_{\mathrm{G}_i \mathrm{G}_j} + C_{\mathrm{I}_i \mathrm{I}_j} +
C_{\mathrm{G}_i \mathrm{I}_j} + C_{\mathrm{I}_i \mathrm{G}_j} , \\
C_{\mathrm{w}_i \mathrm{l}_j} (\ell) &=&
C_{\mathrm{s}_i \mathrm{G}_j} + C_{\mathrm{s}_i \mathrm{I}_j} +
C_{\mathrm{t}_i \mathrm{G}_j} + C_{\mathrm{t}_i \mathrm{I}_j} .
\eeqa

With the Limber's approximation \cite{Limber54,LoVerde08},
we can compute the spectra as
\beq
C_{\mathrm{X}_i \mathrm{Y}_j} (\ell) =
\int_0^{\chi_\mathrm{H}} d\chi \,
\frac{W^\mathrm{X}_i (\chi) W^\mathrm{Y}_j (\chi)}{\chi^2}
P_\mathrm{m} \left( k = \frac{\ell+1/2}{\chi}, \chi \right) ,
\eeq
where $\mathrm{X}, \mathrm{Y} = \{ \mathrm{s}, \mathrm{t}, \mathrm{G}, \mathrm{I} \}$,
kernels $W^\mathrm{X}_i$ are defined in Eqs.~(\ref{eq:Wtkernel}), (\ref{eq:Wskernel}),
(\ref{eq:WGkernel}), and (\ref{eq:WIkernel}),
and $P_\mathrm{m} (k, \chi)$ is the matter power spectrum.
We use linear Boltzmann code \texttt{CAMB} \cite{Lewis00}
to generate transfer function for total matter component.
For our interested scales, the nonlinear evolution of the matter fluctuation is
important. Hence, we employ the \texttt{HALOFIT} scheme \cite{Smith03}
to compute nonlinear matter power spectra adopting parameters in Ref.~\cite{Takahashi12}.

\subsection{Covariance matrix}
For simplicity, we adopt the Gaussian covariance matrix,
\beqa
\mathrm{Cov}[C_\mathrm{UV} (\ell) ,C_\mathrm{XY} (\ell')] =
\frac{4 \pi}{\Omega_s} \frac{\delta_{\ell \ell'}}{(2 \ell + 1) \Delta \ell}
\nonumber \\
\times [ \hat{C}_\mathrm{UX} (\ell) \hat{C}_\mathrm{VY} (\ell) +
\hat{C}_\mathrm{UY} (\ell) \hat{C}_\mathrm{VX} (\ell)] ,
\label{eq:covariance}
\eeqa
where $\Omega_s$ is the area of the survey region,
$\Delta \ell$ is the width of the multipole bins and
the subscripts $\mathrm{U}$, $\mathrm{V}$, $\mathrm{X}$, and $\mathrm{Y}$
denote types of observables and redshift bins, i.e.,
$\mathrm{w}_i \, (i = 1, \ldots, N_\mathrm{w})$
and $\mathrm{l}_i \, (i = 1, \ldots, N_\mathrm{l})$.
The shot noise in GW source number density and shape noise in WL are included as
\beq
\hat{C}_\mathrm{XY} = C_\mathrm{XY} + \delta_\mathrm{XY} N_\mathrm{X} ,
\eeq
where $\delta_\mathrm{XY}$ is the Kronecker delta which takes unity only when
the types of observables and the bins of redshifts are the same and otherwise zero,
and
\beq
N_{\mathrm{w}_i} = \frac{1}{\bar{n}^\mathrm{w}_i}, \,
N_{\mathrm{l}_i} = \frac{\sigma_\gamma^2}{\bar{n}^\mathrm{l}_i} ,
\label{eq:noiseterm}
\eeq
where $\sigma_\gamma$ is the intrinsic variance of galaxy shape
and $\bar{n}^\mathrm{l}_i$ and $\bar{n}^\mathrm{w}_i$ is the number density per steradian
in the $i$th bin for weak lensing source galaxies and GW sources
(Eq.~\ref{eq:meandensity}), respectively.

\section{Results}
\label{sec:results}

\subsection{Surveys}
Here, we characterize surveys for measurements of
auto- and cross-spectra of GW source distributions and weak lensing.

First, we specify survey parameters for GW observation with Einstein Telescope.
Based on the first observing run and first detection of the binary NS event by Advanced LIGO,
the inferred binary BH merger rate density is $9 \text{--} 240 \, \Gpc^{-3} \, \mathrm{yr}^{-1}$ \cite{Abbott16b}
and binary NS merger rate density is $320 \text{--} 4740 \, \Gpc^{-3} \, \mathrm{yr}^{-1}$ \cite{Abbott17d}.
The merger rate density has a possibility to evolve with time \cite{Dominik13}.
For simplicity we assume the event rate density is $\dot{n}_\mathrm{GW} = 5 \times 10^{-6} \, h^3 \, \Mpc^3 \, \mathrm{yr}^{-1}$
regardless of redshifts and the duration of observation is $T_\mathrm{obs} = 1 \, \mathrm{yr}$.
This event rate roughly corresponds to the optimistic estimate of binary NS event
which can be detected by Advanced LIGO.
Accordingly, this detection rate is feasible for Einstein Telescope,
which has much better sensitivity than Advanced LIGO.
For bias parameter, we parametrize it based on Refs.~\cite{Fry96,Tegmark98}, as
\beq
b_\mathrm{GW} (z) = b_\mathrm{w1} + \frac{b_\mathrm{w2}}{D_+ (z)} ,
\label{eq:GWbias}
\eeq
where $b_\mathrm{w1}$ and $b_\mathrm{w2}$ are free parameters and marginalized in the analysis.
For binning of luminosity distances, equivalently redshifts,
we adopt the number of bins as $N_\mathrm{w} = 6$
and equally spaced bins with respect to redshifts in the range of $0.3 < z < 2.7$.

\begin{figure}[htbp]
  \centering
  \includegraphics[width=0.45\textwidth]{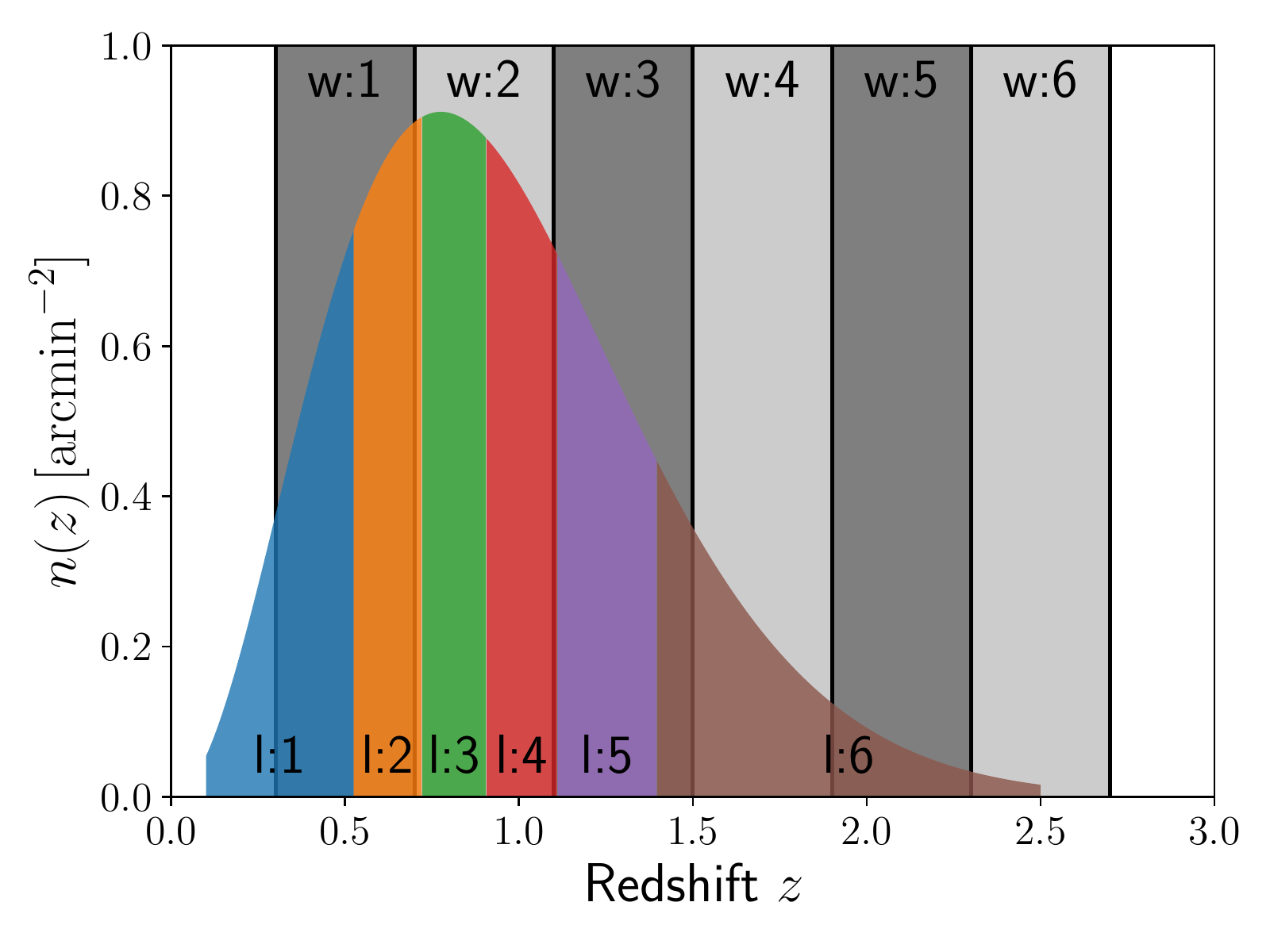}
  \caption{Configuration of redshift bins for weak lensing and GW source distributions.
  The colored (gray) regions correspond to the bins for weak lensing (GW source distribution).}
  \label{fig:zbin}
\end{figure}

Next, let us consider weak lensing surveys.
The survey area of weak lensing with {\it Euclid} is taken as
$\Omega_ s = 15000 \, \mathrm{deg}^2$
and intrinsic variance of galaxy shape is $\sigma_\gamma = 0.22$ \cite{Amendola13}.
Since the resolution of localization of GW sources
is order of $10 \, \mathrm{deg}^2$ \cite{Abbott18}
and the current ground-based surveys span $\sim 100\text{--}1000 \, \mathrm{deg}^2$,
the scales available for cross-correlations are quite limited
when WL measurements with ground-based surveys are employed.
On the other hand, the \textit{Euclid} survey, which covers much larger areas,
has advantage in wide dynamic range of angular scales for cross-correlation measurements.
The functional form of the source number density is given as
\beq
n(z) \propto \left( \frac{z}{z_0} \right)^2
\exp \left[ - \left( \frac{z}{z_0} \right)^{1.5} \right] ,
\label{eq:nofz}
\eeq
where $z_0 = 0.64$, which roughly corresponds to
the mean redshift $z_\mathrm{mean} = 0.9$ \cite{Amendola13}.
This distribution is normalized as
\beq
\int_{z_\mathrm{min}}^{z_\mathrm{max}} n(z) dz = n_0,
\label{eq:nnorm}
\eeq
where $n_0 = 30\, \mathrm{arcmin}^{-2}$ is the total source density,
and the minimum (maximum) redshift is set as $z_\mathrm{min} = 0.1$
($z_\mathrm{max} = 2.5$) \cite{Amendola13}.
Since \textit{Euclid} provides accurate photometric redshift,
we ignore the scatters of photometric redshifts.
Then, the number density in the $i$th lensing bin is given as,
\beq
p_i (z) \propto
  \begin{cases}
    n(z) & (z_{i, \mathrm{min}} < z < z_{i, \mathrm{max}}) \\
    0 & (\text{otherwise}) .
  \end{cases}
\eeq
Note that $p_i (z)$ should be normalized as in Eq.~(\ref{eq:pnorm})
and $p_i (z) dz = p_i (\chi) d\chi$.
Here, we consider six lensing bins ($N_\mathrm{l} = 6$).
We determine the bin configuration so that in each bin the number density
of source galaxies becomes the same.
Figure~\ref{fig:zbin} and Table~\ref{tab:binning} show
the binnings of GW source distribution and weak lensing.

\begin{table}
\caption{Redshift binning.}
\begin{ruledtabular}
\begin{tabular}{ccc}
  Bin & GW source distribution & Weak lensing \\
  \hline
  1 & $0.3 < z < 0.7$ & $0.10 < z < 0.52$ \\
  2 & $0.7 < z < 1.1$ & $0.52 < z < 0.72$ \\
  3 & $1.1 < z < 1.5$ & $0.72 < z < 0.90$ \\
  4 & $1.5 < z < 1.9$ & $0.90 < z < 1.11$ \\
  5 & $1.9 < z < 2.3$ & $1.11 < z < 1.39$ \\
  6 & $2.3 < z < 2.7$ & $1.39 < z < 2.50$ \\
\end{tabular}
\end{ruledtabular}
\label{tab:binning}
\end{table}

Finally, let us define the binning of multipoles for auto- and cross-spectra.
We fix the minimum multipole as $\ell_\mathrm{min} = 10$ and
consider two different cases for maximum multipoles, $\ell_\mathrm{max} = 100, \, 300$.
With the interferometer network of Advanced LIGO, Virgo, KAGRA, and LIGO-India,
the median of localization at 95\% confidence level
is $9\text{--}12 \, \mathrm{deg}^2$ \cite{Abbott18},
which corresponds to the multipole of $\sim 100$.
Thus, in the era of Einstein Telescope,
even maximum multipole of $\ell_\mathrm{max} = 300$ is expected to be possible.
The bins are logarithmically equally spaced and the number of bins is $30$.
We summarize parameters which characterize the surveys in Table~\ref{tab:values}.

\begin{table*}
\caption{Summary of parameters}
\begin{ruledtabular}
\begin{tabular}{cclc}
  \multicolumn{4}{c}{Fixed parameters} \\
  Symbol & Value & Explanation & Reference \\
  \hline
  $\sigma_{\ln D}$ & $0.05$ & Standard deviation of the luminosity distance distribution. & Eq.~(\ref{eq:xD}) \\
  $T_\mathrm{obs}$ & $1\, \mathrm{yr}$ & Duration of GW observation. & Eq.~(\ref{eq:meandensity}) \\
  $\dot{n}_\mathrm{GW}$ & $5 \times 10^{-6} \, h^3 \, \Mpc^{-3} \, \mathrm{yr}^{-1}$ & Mean number density of GW events per unit time. & Eq.~(\ref{eq:meandensity}) \\
  $\Omega_s$ & $15000 \, \mathrm{deg}^2$ & Area of the survey region. & Eq.~(\ref{eq:covariance}) \\
  $z_0$ & $0.64$ & Redshift parameter of lensing source distribution. & Eq.~(\ref{eq:nofz}) \\
  $n_0$ & $30 \, \mathrm{arcmin}^{-2}$ & Lensing source number density. & Eq.~(\ref{eq:nnorm}) \\
  $\sigma_\gamma$ & $0.22$ & Intrinsic variance of shapes of source galaxies. & Eq.~(\ref{eq:noiseterm}) \\
  $C_1$ & $5 \times 10^{-14} \, h^{-2} \, \mathrm{M}_\odot^{-1} \, \Mpc^3$ & Normalization of intrinsic alignment. & Eq.~(\ref{eq:WIkernel}) \\
  \hline
  \multicolumn{4}{c}{Varied parameters} \\
  Symbol & Fiducial value & Explanation & Reference \\
  \hline
  $b_\mathrm{w1}$, $b_\mathrm{w2}$ & $1$, $1$ & Bias parameters for GW source number density distribution. & Eq.~(\ref{eq:GWbias}) \\
  $A_\mathrm{IA}$ & $1$ & Amplitude of intrinsic alignment. & Eq.~(\ref{eq:WIkernel}) \\
  $\Omega_\mathrm{m}$ & $0.3153$ & Matter density at the present Universe normalized by critical density. & \\
  $h$ & $0.6727$ & Hubble parameter in the unit of $100 \, \mathrm{km}/\mathrm{s}/\Mpc$. & \\
  $w_\mathrm{de}$ & $-1$ & Equation of state parameter of dark energy. & \\
  $\sigma_8$ & $0.831$ & The amplitude of matter fluctuation at the scale of $8 \, \hMpc$. & \\
\end{tabular}
\end{ruledtabular}
\label{tab:values}
\end{table*}

\subsection{Spectra with fiducial parameters}
\begin{figure*}[htbp]
  \centering
  \includegraphics[width=0.9\textwidth]{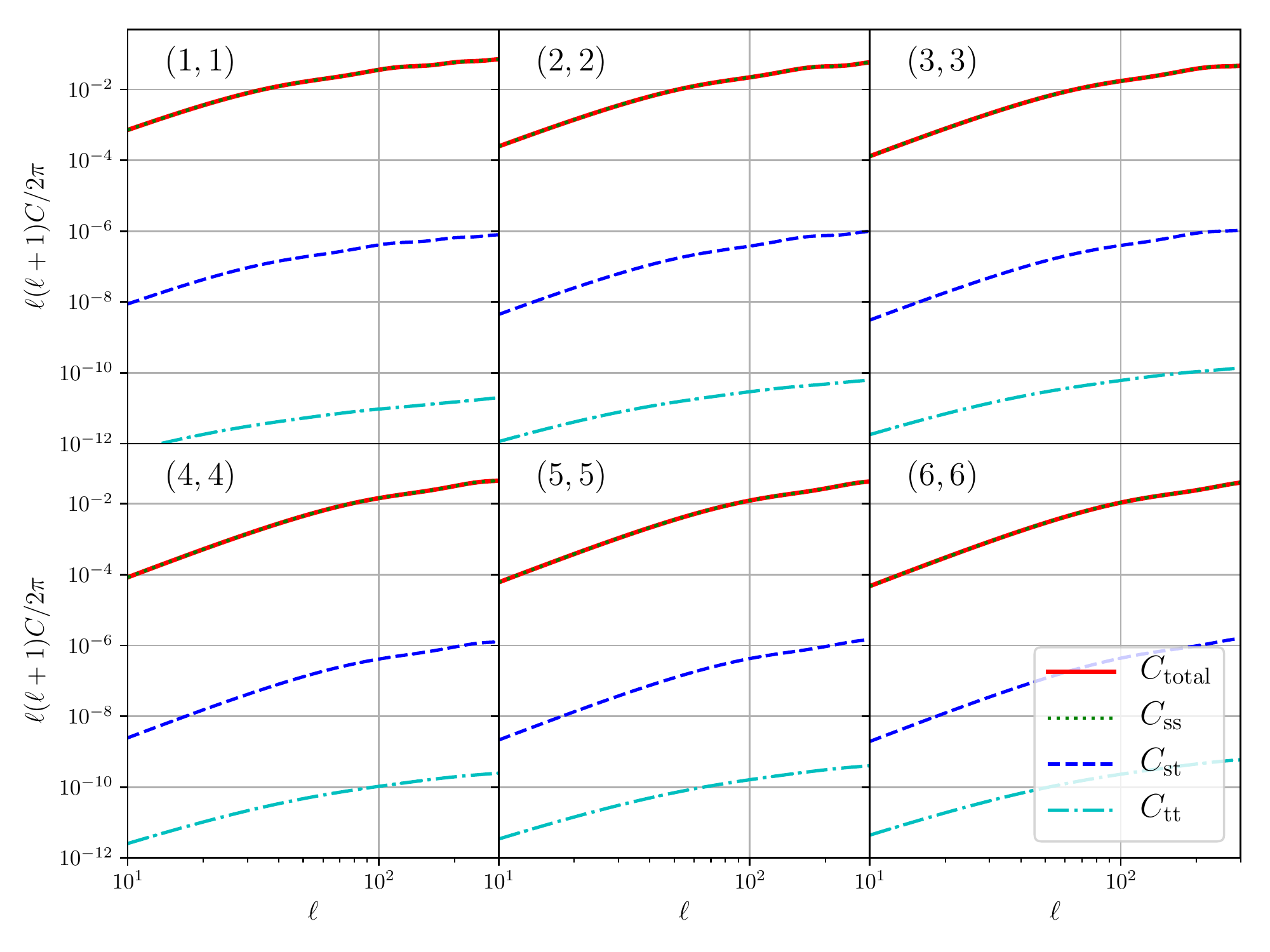}
  \caption{The auto-power spectra of GW source distributions.
  The numbers in parenthesis denote the bins.}
  \label{fig:wwpower}
\end{figure*}

\begin{figure*}[htbp]
  \centering
  \includegraphics[width=0.9\textwidth]{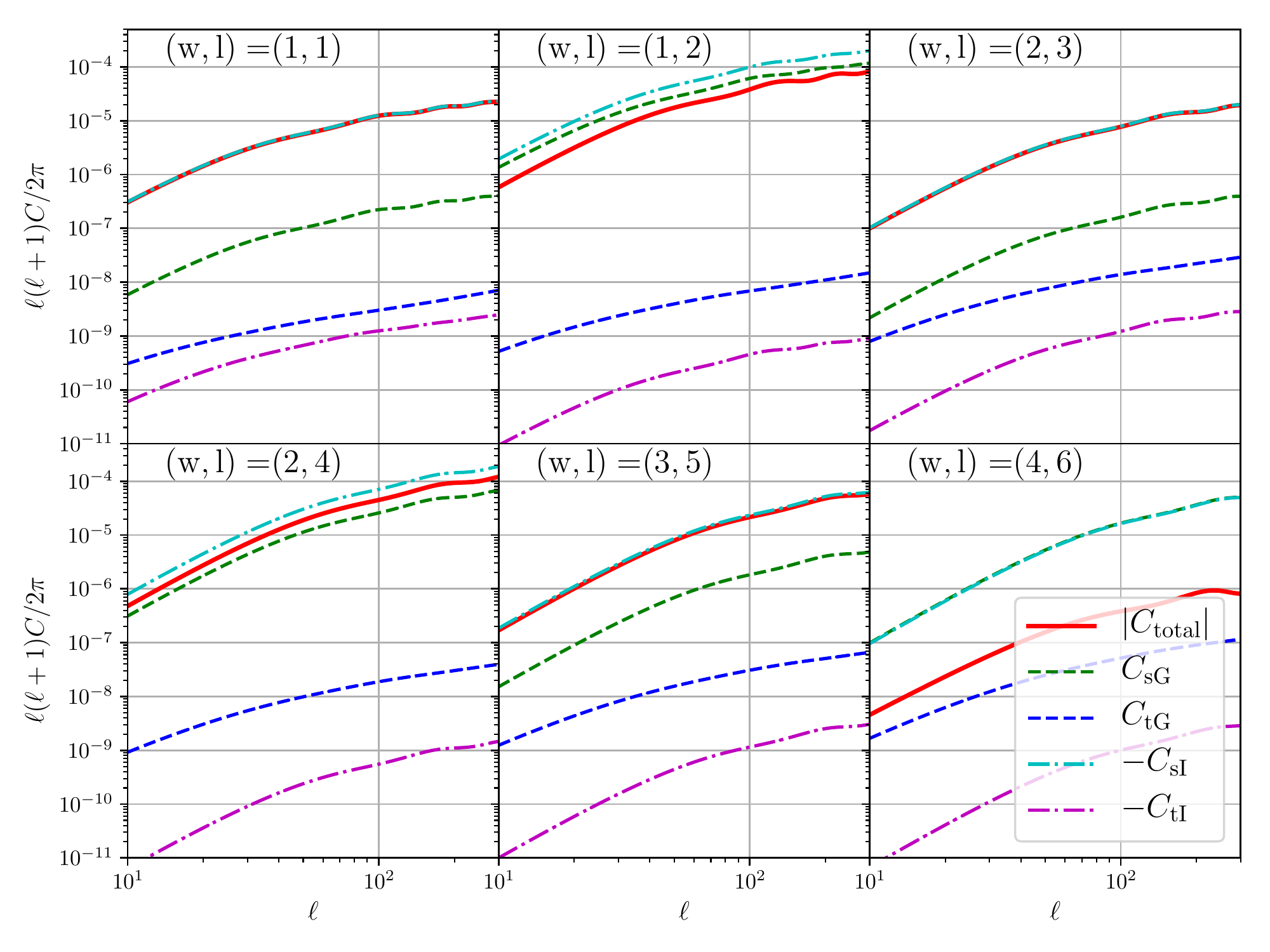}
  \caption{The cross-power spectra of GW source distributions and tomographic weak lensing.
  The numbers in parenthesis denote the bins.
  Note that cross-correlations with IA term is always negative.
  Since the total spectra can be positive or negative, we show the absolute values for the spectra.}
  \label{fig:wlpower}
\end{figure*}

\begin{figure*}[htbp]
  \centering
  \includegraphics[width=0.9\textwidth]{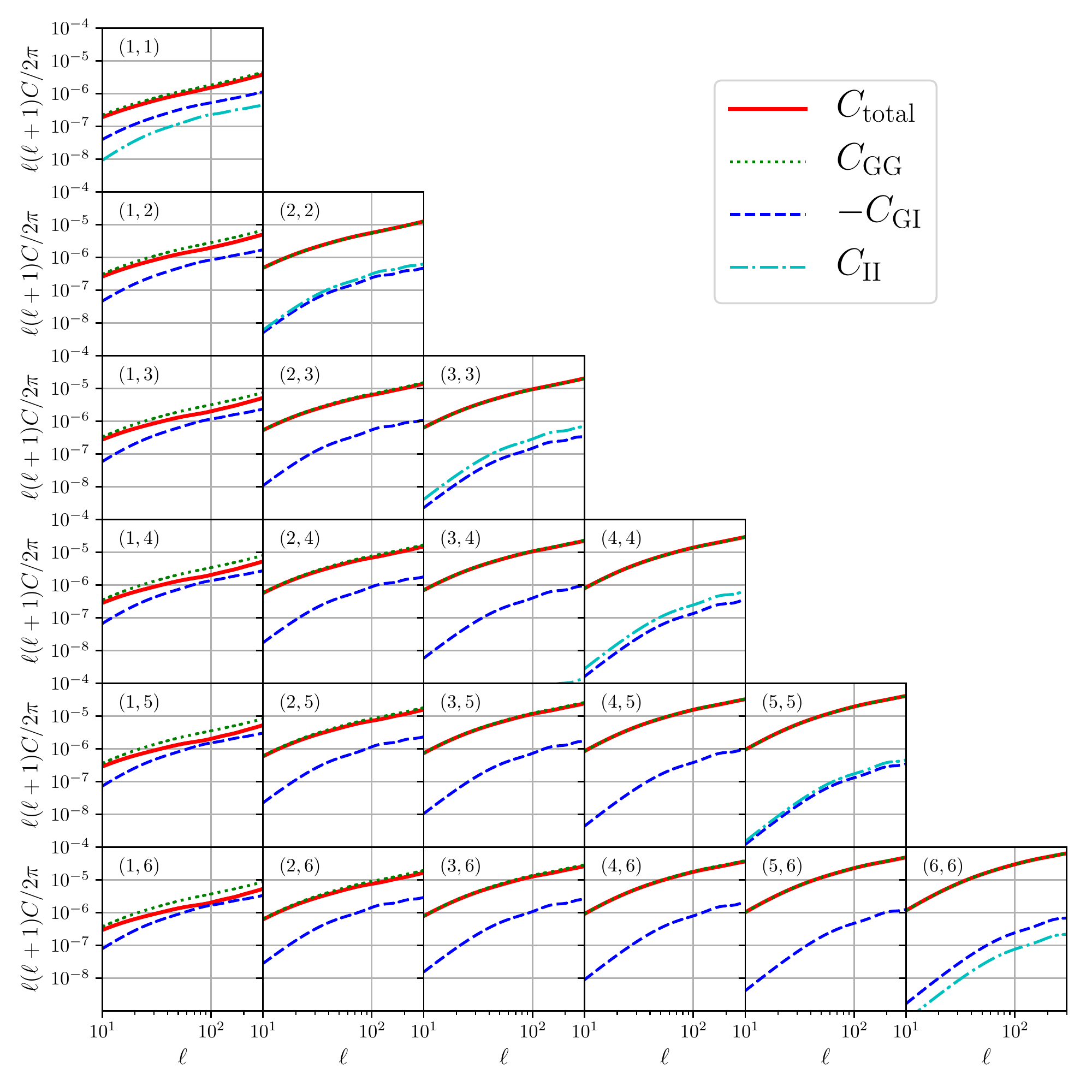}
  \caption{The auto-power spectra of tomographic weak lensing.
  The numbers in parenthesis denote the bins.
  Note that cross-correlations between lensing and IA are always negative.}
  \label{fig:llpower}
\end{figure*}

In Figures~\ref{fig:wwpower}, \ref{fig:wlpower}, and \ref{fig:llpower},
auto- and cross-power spectra are shown.
We compute these spectra with fiducial parameters listed in Table~\ref{tab:values}.
For weak lensing, we can cross-correlate $N_\mathrm{l} (N_\mathrm{l}+1)/2 = 21$ pairs
of lensing bins and all of them have appreciable signals.
Though we can take cross-correlation for $N_\mathrm{w} (N_\mathrm{w}+1)/2 = 21$ pairs
for GW source distributions, correlation between different bins is suppressed
because the deviation of luminosity distance from true one is assumed to be small
in Eq.~(\ref{eq:xD}).
Therefore auto-correlations contain most of information for GW source distributions.
For cross-spectra between GW source distribution and weak lensing,
there are $N_\mathrm{l} \times N_\mathrm{w} = 36$ spectra.
In total there are $78$ spectra used in the analysis.
In Figure~\ref{fig:wlpower}, we show spectra where the redshift ranges
of two bins are overlapped. In this case, the contribution due to IA
is appreciable because the support of IA kernel is confined
contrast to wide support of lensing kernel.
When GW source distribution bin is located farther than lensing bin,
the resultant spectrum is close to zero.

\subsection{Fisher forecast}
\begin{figure}[htbp]
  \centering
  \includegraphics[width=0.45\textwidth]{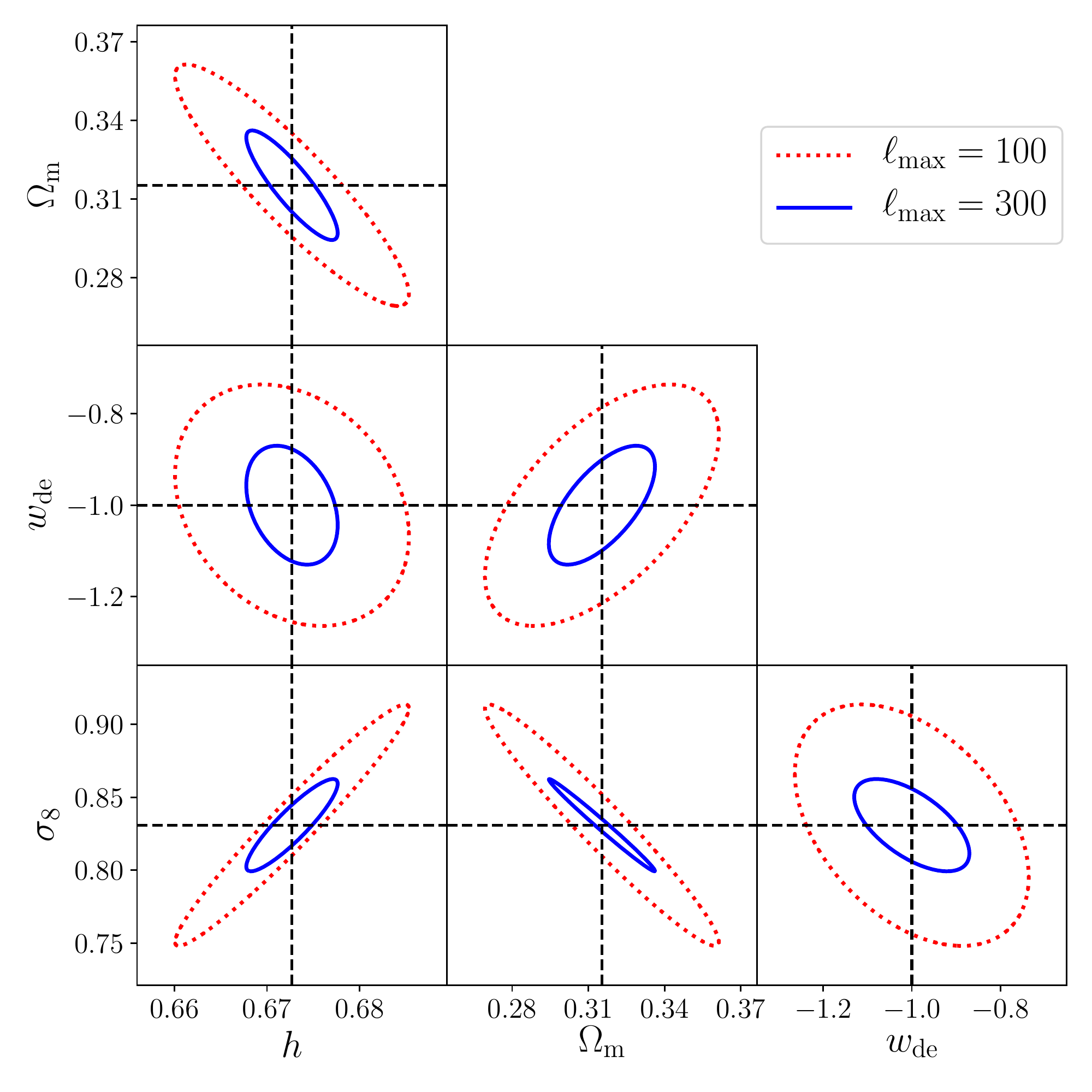}
  \caption{Projected confidence regions at $68\%$ level of cosmological parameters
  $(h, \Omega_\mathrm{m}, w_\mathrm{de}, \sigma_8)$
  from the Fisher matrix. The red dashed (blue solid) line corresponds to the result
  with the maxmimum multipole $\ell_\mathrm{max} = 100$ ($\ell_\mathrm{max} = 300$).
  The black dashed lines show fiducial values.}
  \label{fig:region}
\end{figure}

In this Section, we present forecast of parameter constraints based on
Fisher matrix approach \cite{Tegmark97}.
Since we assume that the covariance matrix does not depend on parameters
and there are no correlations between different multipoles,
the Fisher matrix can be simplified as
\beq
F_{\alpha \beta} =
\sum_\ell \sum_{\mathrm{U}, \mathrm{V}, \mathrm{X}, \mathrm{Y}}
\frac{\partial C_\mathrm{UV} (\ell)}{\partial p_\alpha}
\mathrm{Cov}[C_\mathrm{UV} (\ell), C_\mathrm{XY} (\ell)]^{-1}
\frac{\partial C_\mathrm{XY} (\ell)}{\partial p_\beta} ,
\eeq
where $p_\alpha$ denotes a cosmological or nuisance parameter.
The marginalized error for the parameter $p_\alpha$ is given as
\beq
\sigma(p_\alpha) = \sqrt{(F^{-1})_{\alpha \alpha}} .
\eeq
We consider the parameter space of
$(h, \Omega_\mathrm{m}, w_\mathrm{de}, \sigma_8, b_\mathrm{w1}, b_\mathrm{w2}, A_\mathrm{IA})$,
where the first four parameters are our interested cosmological parameters and
the latter three are nuisance parameters.
When varying matter density $\Omega_\mathrm{m}$,
we fix baryon density $\Omega_\mathrm{b}$ and vary only cold dark matter density $\Omega_\mathrm{c}$.
For nuisance parameters, we always marginalize them in this analysis.
We show marginalized errors for cosmological parameters in Table~\ref{tab:errors}
and projected 68$\%$ level confidence regions with
auto- and cross-spectra between GW distributions and weak lensing
for two different cases of maximum multipoles $\ell_\mathrm{max} = 100, \, 300$
in Figure~\ref{fig:region}.
The results show one can place a tight constraint on cosmological parameters
with three different types of spectra.
Especially, in addition to the dark energy parameter $w_\mathrm{de}$,
we can constrain the amplitude of matter fluctuation $\sigma_8$,
which is degenerate with galaxy bias when galaxy clustering measurement is used.

Recently, it has been reported that there is a tension between estimates of
Hubble parameter $H_0$ from type Ia supernovae (SNe Ia) observations and CMB measurements.
SNe Ia observations measure the distance-redshift relation in the nearby ($z < 1$) Universe.
On the other hand, measurements of CMB probe into
the distance scale in distant ($z > 1000$) Universe
with acoustic patterns in angular power spectrum.
Therefore, the tension may imply deviation from the standard cosmological model.
In order to confirm existence of the tension,
precise measurement of Hubble parameter is critical.
The current estimates of Hubble parameter are
$H_0 = 67.27 \pm 0.66 \, \mathrm{km} \, \mathrm{s}^{-1} \, \mathrm{Mpc}^{-1}$
for CMB measurements of the \textit{Planck} mission (TT,TE,EE+lowP) \cite{Planck16} and
$H_0 = 73.24 \pm 1.74 \, \mathrm{km} \, \mathrm{s}^{-1} \, \mathrm{Mpc}^{-1}$
for SNe Ia observations \cite{Riess16}.
Our forecasted precision of Hubble parameter is
$\sigma (H_0) = 0.33 \, \mathrm{km} \, \mathrm{s}^{-1} \, \mathrm{Mpc}^{-1}$
with $\ell_\mathrm{max} = 300$.
As a result, with the auto- and cross-correlations of GW source distributions and WL,
the above discrepancy can be distinguished at $18\text{-}\sigma$ significance level.
Furthermore, these correlations will provide independent estimates from large-scale structures
at intermediate redshifts ($z \sim 1\text{--}2$).
Thus, the correlations can be a promising and powerful
probe into the distance-redshift relation in the coming era.

\begin{table}
\caption{Marginalized errors from Fisher matrix.}
\begin{ruledtabular}
\begin{tabular}{ccccc}
  Maximum multipole & $\sigma (h)$ & $\sigma (\Omega_\mathrm{m})$ & $\sigma (w_\mathrm{de})$ & $\sigma (\sigma_8)$\\
  \hline
  $\ell_\mathrm{max} = 100$ & $0.0084$ & $0.031$ & $0.17$ & $0.055$\\
  $\ell_\mathrm{max} = 300$ & $0.0033$ & $0.014$ & $0.086$ & $0.021$\\
\end{tabular}
\end{ruledtabular}
\label{tab:errors}
\end{table}

\section{Conclusions}
\label{sec:conclusions}
The discovery of GW signals from BH binary merger by Advanced LIGO
has opened a new window into astrophysics and cosmology.
From the observed wave forms, we can infer the absolute luminosity
of GW and then measure the luminosity distance of the sources.
If the redshifts of the sources are available,
we can probe into the geometry of the Universe via the distance-redshift relation.
Although it has already been reported that
the source redshift is identified from the EM counterpart
for the NS binary merger event GW170817, measuring the source redshift is still challenging
especially for BH binary merger.
However, without redshift information, we can explore the distance-redshift relation
by combining another observable which redshift information is accessible.

In this work, we focus on cross-correlating weak gravitational lensing
with the number density distributions of GW sources.
WL is an unbiased tracer of matter distribution in the Universe
and one of main observational targets for upcoming imaging surveys.
We employ tomographic technique, where
the whole source galaxy samples are divided according to their photometric redshifts.
Thus we can efficiently extract information of the large-scale structures
in different redshifts.
We show that auto- and cross-correlations of GW source distributions and WL
enable us to obtain tight constraints on cosmological parameters
based on Fisher matrix approach in the case with \textit{Euclid} for WL
and Einstein Telescope for GW source distributions.
One of advantages of using WL over galaxy clustering is that
galaxy bias is not necessary and we can constrain the amplitude of
matter power spectrum, which is degenerate with galaxy bias.
Thus we can place a tight constraint without being degraded by nuisance parameters
like galaxy bias.
Furthermore, the tight constraint on Hubble parameter has a possibility
to reconcile the tension between SNe Ia observations and CMB measurements.

Finally, we would like to discuss future prospects for standard sirens.
Recently, several works present predictions of angular power spectrum
of GW energy distribution \cite{Cusin18a,Cusin18b}.
Though auto-spectra of GW energy distribution contain information about cosmology
and astrophysics, by combining with other observables such as WL,
we can obtain more information and evade systematic effects like intrinsic alignments.
Another topic which should be addressed is three dimensional correlations
of GW source distributions.
In this work, we focused only on projected quantities.
Since projection mixes Fourier modes of small and large scales,
we can efficiently obtain independent information from three dimensional correlations.
There is a possibility that three dimensional clustering of GW sources and
cross-correlation between GW source distributions and other observables,
e.g., the spatial distribution of spectroscopically detected galaxies,
can enable us to probe into the geometry of the Universe.
We leave it for future work.


\begin{acknowledgments}
The author thanks Masamune Oguri and Kentaro Komori for helpful discussions.
KO is supported by Research Fellowships of the Japan Society for the
Promotion of Science (JSPS) for Young Scientists,
and Advanced Leading Graduate Course for Photon Science.
This work was supported by JSPS Grant-in-Aid for JSPS Research Fellow
Grant Number JP16J01512.
\end{acknowledgments}

\bibliographystyle{apsrev4-1}
\bibliography{main}

\end{document}